# Visualizing buried local carrier diffusion in halide perovskite crystals via two-photon microscopy

*Camille Stavrakas§, Géraud Delport §, Ayan A. Zhumekenov, Miguel Anaya, Rosemonde Chahbazian, Osman M. Bakr, Edward S. Barnard and Samuel D. Stranks\**

Dr. Camille Stavrakas, Dr. Géraud Delport, Dr. Miguel Anaya, Rosemonde Chahbazian, Dr. Samuel D. Stranks, Cavendish Laboratory, JJ Thomson Avenue, Cambridge CB3 0HE, UK.
\*E-mail: sds65@cam.ac.uk

Ayan A. Zhumekenov, Prof. Osman M. Bakr, Division of Physical Sciences and Engineering, King Abdullah University of Science and Technology (KAUST), Thuwal 23955-6900, Kingdom of Saudi Arabia

Dr. Edward S. Barnard,
Molecular Foundry, Lawrence Berkeley National Laboratory, Berkeley, CA, USA

Halide perovskites have shown great potential for light emission and photovoltaic applications due to their remarkable electronic properties and compatibility with cost-effective fabrication techniques. Although the device performances are promising, they are still limited by microscale heterogeneities in their photophysical properties. In particular, the relation between local heterogeneities and the diffusion of charge carriers at the surface and in the bulk, crucial for efficient collection of charges in a light harvesting device, is not well understood. Here, a photoluminescence tomography technique is developed in a confocal microscope using one- and two-photon excitation to distinguish between local surface and bulk diffusion of charge carriers in methylammonium lead bromide single crystals. The local temporal diffusion is probed at various excitation depths to build statistics of local electronic diffusion coefficients. The measured values range between 0.3 to 2 $cm^2.s^{-1}$ depending on the local trap density and the morphological environment – a distribution that would be missed from analogous macroscopic or surface-measurements. Tomographic images of carrier diffusion were reconstructed to reveal buried crystal defects that act as barriers to carrier transport. This work reveals a new framework to understand and homogenise diffusion pathways, which are extremely sensitive to local properties and buried defects.

Over the past ten years, halide perovskites have emerged as strong candidates for various light-harvesting and light-emission applications[1–3]. The performances of perovskite-based photovoltaics (PV) and light-emitting diodes (LEDs) are now competing with mature, commercial technologies[4]. This rapid development has been made possible by the design of new halide perovskite compositions[5–7] which generally share properties of remarkably long carrier diffusion lengths (0.1-1 µm)[8,9] even when simple cost-effective fabrication techniques are employed. However, for halide perovskites to reach their full potential, one has to understand the microscopic heterogeneities that still limit their performances[10,11]. For instance, local defects, both at the surface and inside the bulk, trap charge carriers thus limiting their ability to diffuse through the material. It is therefore critical to investigate the diffusion mechanisms at the local scale to identify these trap sites and elucidate ways to mitigate their influence on carrier diffusion and recombination.

Methylammonium lead bromide ($MAPbBr_3$, MA=$CH_3NH_3^+$) single crystals have remarkable photophysical properties as highlighted in recent reports on amplified spontaneous emission[12] and lasing phenomena[13,14], two-photon absorption[15,16], extreme sensitivity to environment[17], excitonic properties[18,19], and long carrier diffusion lengths[20]. Additionally, their optical properties are well-documented including their refractive index[21,22] and exciton binding energy[23], and photon reabsorption has been quantified[22,24,25]. Such single crystals are ideal platforms to investigate intrinsic charge carrier recombination and transport because they will not be as influenced by morphological properties as their polycrystalline film counterparts, where grain boundaries may have a dominant impact on transport[26,27]. On one hand, the surface properties of these single crystals, such as defect densities[17] and carrier diffusion, have been reported[28,29]. On the other hand, optoelectronic properties are more difficult to probe within the bulk of these crystals, particularly on the microscale, due to the large optical absorption coefficients of these materials[22]. Time-resolved PL (TRPL) microscopy measurements allow us to study diffusive effects on the micro-scale[9,29,30]. Most TRPL studies on halide perovskites to date are based on one photon (1P) excitation techniques[8,31] which, due to the short optical absorption depth in halide perovskites[22,24], typically probes the top ~50-100 nm of the sample with most commonly used visible excitation wavelengths. These techniques are therefore particularly sensitive to effects which are most prominent on the surface[32–34] that include surface defects[35], light soaking[30], waveguiding[36] and surface irregularities[37]. Therefore, it is not possible to observe the diffusion of charge carriers deeper in the crystal using a 1P technique. Furthermore, many studies deduce diffusion properties[8,38,39] from macroscopic 1P TRPL measurements, missing crucial local variations in carrier lifetime and diffusion properties that are ultimately responsible for power losses in devices.

Recently, we combined 1P and two-photon (2P) TRPL confocal microscopy with excitation and emission fixed at the same spatial location to unveil local, buried carrier recombination sites in halide perovskites that cannot be observed through 1P measurements alone[40]. Here, we further adapt a 1P/2P TRPL confocal microscope setup to collect the photons emitted at locations at a controllable distance away from the excitation area using a scanning collection setup[41]. By performing these diffusion measurements as a function of depth on MAPbBr$_3$ single crystals, we determine the diffusion properties in the bulk of the crystals and compare these findings with their surface diffusion properties. We use this technique to reveal a spatially and depth-dependent heterogeneous distribution of carrier diffusion properties. We then construct time and spatially resolved images of carrier diffusion and use these images to visualise buried crystal defects that have an impact on carrier transport. These results give critical insight into the factors that limit carrier transport in halide perovskite materials.

In Figure 1a, we show a general schematic of our experimental setup to probe carrier diffusion in four dimensions (time and 3D space). In general, we adjust the depth at which we generate photo-excited carriers (and probe diffusion) by using either 1P excitation ($z=0$) or 2P excitation ($z>0$). At a given depth, we measure a series of TRPL decay curves at different positions at distance *x* away from the fixed excitation spot (at *x*=0) by raster scanning the emission collection (Figure 1b, see Experimental section and Supporting Information (SI) for details). In Figure 1c, we show a schematic representing the impact of the carrier diffusion on the width of the PL spatial distribution, characterized by the standard deviation $\sigma_x$ of a Gaussian PL profile.

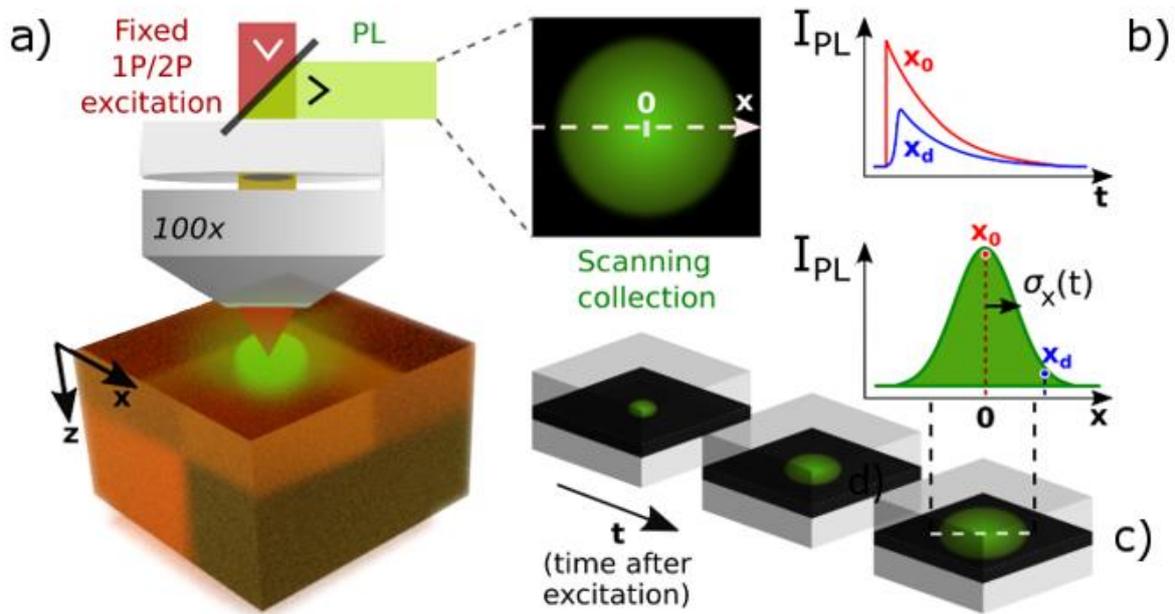

*Figure 1: Overview of the time and spatially resolved PL microscope setup for measuring local carrier diffusion. a) Schematic of the TRPL experimental setup (1P or 2P) to probe the diffusion properties laterally at different distance (x) from the excitation spot. b) Representation of the TRPL decays that can be measured with this setup, shown here for two different x positions: $x_0$ (center, i.e. x=0) and $x_d$ (away from the center). c) Artistic view of the impact of the diffusion of carriers leading to a broadening of the spatial distribution of the PL with time, including the definition of the standard deviation $\sigma_x$ associated with the Gaussian distributions employed in this work.*

We grew MAPbBr$_3$ single crystals using an Inverse Temperature Crystallization method[42,43] (see Experimental Section). We show in Figure 2a a series of example decay curves for 1P excitation ($z$=0) in a crystal at distance $x$ away from the local excitation spot ($x$=0) (see Fig. S3 for the full series of PL decays). We use an excitation wavelength of 405 nm and fluence of 1.3 µJ.cm$^{-2}$, which generates local excitation charge-carrier densities on order ~10$^{17}$ cm$^{-3}$ (see SI for details); the PL emission peak in these samples is at ~540 nm[22,24]. From these decay curves, we determine the PL intensity $I_{PL}(x,t)$ corresponding to each position $x$ and time $t$ after excitation. We see in Figure 2a that the $I_{PL}$ values decrease with $x$ as we move away from the excitation centre at $x$=0. From the TRPL curves, we can select a given time snapshot $t$ and reconstruct the spatial profile $I_{PL}(x,t)$ of the emitted photons over the horizontal $x$ axis (see dotted line in Figure 2a). In Figure 2b, we show the evolution of the extracted spatial distributions in $x$ at selected time snapshots after the initial excitation (t=0) at $x$=0 (see Figure S2 for a larger series). This spatial distribution broadens as a function of time as carriers transport away from the excitation spot.

To characterise the diffusion, we apply a Gaussian fit to the PL profiles at different time delays. This allows us to extract the standard deviation $\sigma_x(t)$ that can be interpreted as the instantaneous diffusion length at time $t$ (see Figure 2b). In Figure 2c we show these standard deviations as a function of time after excitation obtained from the Gaussian fits; we do this separately for the right (x>0) and left (x<0) sides of the excitation spot to characterise any differences in diffusion properties in each region of the crystal. The initial value of $\sigma_x \simeq 440$ nm at $t$=0 originates from a combination of factors, including the optical resolution of the setup ($\sigma_{resol} \simeq 180$ nm in excitation at 405 nm and $\simeq 240$ nm in emission at 540 nm, see SI for details) and the possibility of early time diffusion or reabsorbed photons emitted at early times within the temporal instrument response of the setup ($\simeq 100$ ps).

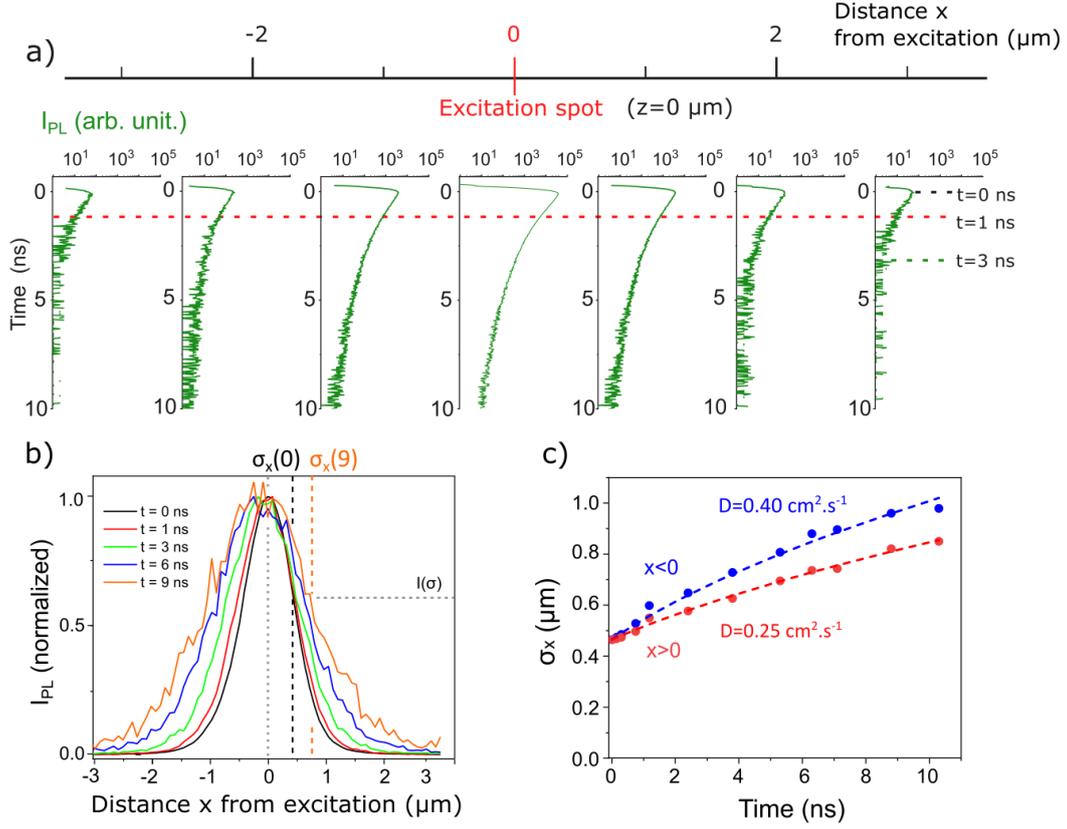

*Figure 2: Surface diffusion properties in MAPbBr$_3$ single crystals. (a) TRPL decay curves at selected collection positions x with 405-nm (1P) excitation at x=0, t=0 (repetition rate of 10 MHz and fluence of 1.3 µJ.cm$^{-2}$). From these data, we extract the normalized PL intensity profiles I$_{PL}$ as a function of time, overlaid in b. The standard deviation σ$_x$(t) extracted from Gaussian fits to the data at each time snapshot t, and the corresponding PL intensity I(σ), are also highlighted in b. (c) Evolution of the σ$_x$ profile broadening as a function of time extracted from the Gaussian TRPL PL diffusion profiles for carriers travelling to the left (x<0, blue) and to the right (x>0, red) of the excitation pulse. Dashed lines indicate fits to the data using Equation (1) that were used to extract the diffusion coefficient values (D) stated in the panel.*

In a classical diffusive scenario, the quantity σ$_x$(t) follows the form[41]:

$$\sigma^2_x(t) = \sigma^2_x(0) + 2Dt \quad (1)$$

where *D* is the carrier diffusion coefficient (see SI for derivation). We find that the evolution of σ is well-fitted by this expression in both regions (dashed lines in Figure 2c). From these fits, we obtain a diffusion coefficient of $D=0.40$ cm$^2$.s$^{-1}$ for the $x<0$ region and $D=0.25$ cm$^2$.s$^{-1}$ for the $x>0$ region. These two values are significantly different, showing that charge carriers diffuse more efficiently on one side than on the other, in line with local heterogeneity in optoelectronic properties in halide perovskites[11,40]. This spatial asymmetry in the diffusion coefficient is also seen in the PL profiles in Figure 2b, which becomes increasingly asymmetric about $x=0$ with time. The measured diffusion coefficients are lower, but of the same order of magnitude, to previously reported values on similar crystals ($\simeq 1$ cm$^2$.s$^{-1}$ [44]). We note that we obtain a higher diffusion coefficient of $D=0.57$ cm$^2$.s$^{-1}$ on another region of the same crystal (see Figure S2), further highlighting the spatial variation of the diffusion properties and the need for microscopic techniques to visualise such variations.

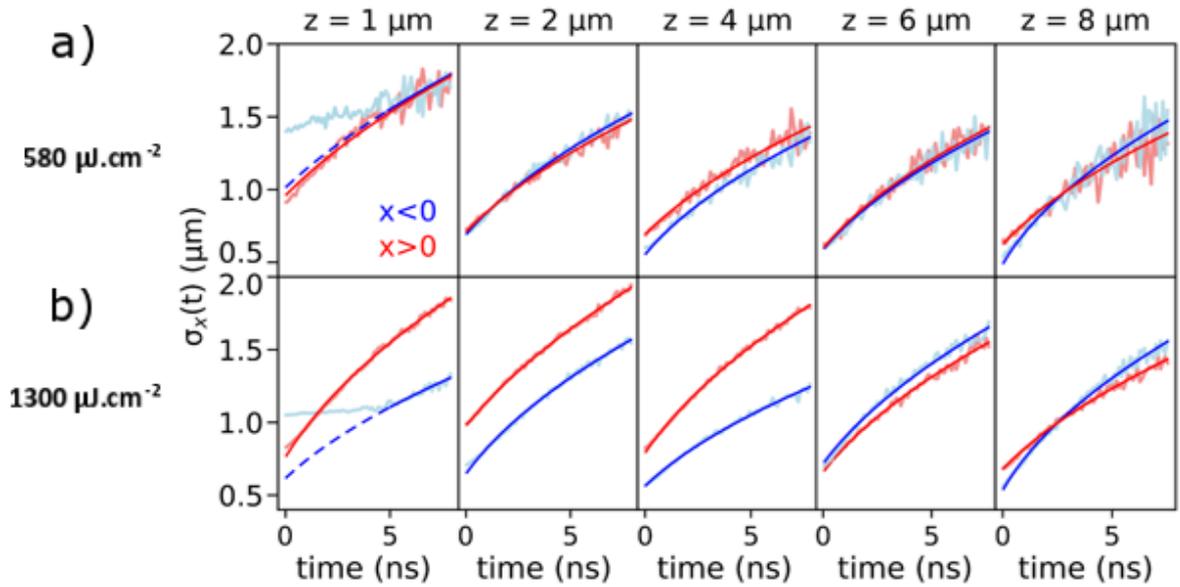

*Figure 3: Bulk diffusion properties in MAPbBr$_3$ single crystals at different depths and fluences. Evolution of the $\sigma_x$ profile broadening as a function of time extracted from the Gaussian TRPL diffusion profiles for x<0 (blue) and x>0 (red) at different depths (z) ascertained using 2P excitation (1200 nm, 8MHz repetition rate) at a fluence of a) 580 µJ.cm$^{-2}$ and b) 1300 µJ.cm$^{-2}$. Solid lines are fits to the data using Eq. 1, with dashed lines indicating extrapolations; the extracted values are plotted in Figure 4.*

After elucidating the local surface diffusion properties ($z=0$) using 1P excitation, we now seek to understand the diffusion properties in the bulk of a MAPbBr$_3$ crystal by selectively exciting at a particular depth ($z>0$) using 2P excitation (1200-nm wavelength). For this purpose, we have used 2P excitation (1200-nm wavelength) to probe a different area of a MAPbBr$_3$ crystal at selected depth ($z>0$). In this configuration, our excitation depth resolution is $\simeq 1.5$ μm and our lateral resolution is $\sigma_{laser} \simeq 0.5$ μm (see SI for details). We show 2P diffusion profiles as a function of depth $z$ in Figure 3a with a 2P fluence of 580 μJ.cm$^{-2}$, which generates a comparable charge excitation density in the samples to the 1P measurements (i.e. ~10$^{17}$ cm$^{-3}$; see SI for details). For each depth, we once again separately treat the regions to the left ($x<0$) and the right ($x>0$). Near the surface at $z=1$ μm, we observe a relatively broad initial PL distribution, $\sigma_x(0)$, for the left ($x<0$) region, which stays constant over several nanoseconds, before showing the classical diffusion dependence of Eq. 1 at later times. We attribute this observation over the first few nanoseconds to be a result of a light soaking effect on the surface due to the extended time required for the 2P measurements, with the local extent of this effect depending on the local PL heterogeneity[30,45]; we note that we also observe this effect in 1P excitation when illuminating for extended times (Figure S4 for further details). By contrast, the temporal evolution of $\sigma_x(t)$ deeper into the crystal, where light soaking effects are far less apparent[40], fits well to the classical diffusion square root law (Eq. 1) across all times (see also Figure S9) and we obtain similar diffusion properties in both the left ($x<0$) and right ($x>0$) regions. We show the depth-dependent diffusion coefficients in Figure 4a, revealing relatively homogeneous values ranging between 0.9 and 1.6 cm$^2$.s$^{-1}$ for $x<0$ and $x>0$ (see statistical distributions in Figure 4c at all depths and regions). These values are notably higher than the values obtained at the surface ($\simeq 0.3$ cm$^2$.s$^{-1}$) and match the highest diffusion coefficients reported from 1P TRPL measurements on MAPbBr$_3$ crystals[29]. The larger values of diffusion coefficient in the bulk than the surface are consistent with the majority of traps residing at the surface, which may limit carrier diffusion in that region[46,47].

To investigate these observations further, we show in Figure 3b the temporal evolution of $\sigma_x(t)$ with higher photo-excitation density (1300 µJ.cm$^{-2}$) and the corresponding extracted depth-dependent diffusion coefficients in Figure 4b. We see a striking increase in the diffusion coefficients at a range of depths particularly for the left ($x<0$) region when compared to the lower fluence measurements. For some depth profiles, the values now reach 2 cm$^2$.s$^{-1}$, thus even exceeding previously reported values[29]. Along with the global increase, we observe a wider distribution of diffusion coefficient values (see Figure 4d). We note that as the fluence increases and the diffusion coefficients generally increase, the measured PL decay times globally decreases from around ≃6 ns to less than 4 ns (see Figure S6) for most of the PL profiles. We attribute these combined observations to a larger saturation of traps at higher fluences[40,48,49], leading to a more efficient diffusion of charge carriers and increased bimolecular recombination (as seen from the shorter PL lifetimes at higher fluence[48]). We note, however, that this saturation of traps is not uniform across all regions, with the diffusion coefficients at some depths remaining relatively unchanged at ≃1 cm$^2$.s$^{-1}$ at higher fluence. This observation suggests that there are heterogeneous distributions of trap densities and perhaps even variations in types of traps below the surface. These local variations in diffusion coefficient laterally and with depth would be missed using macroscopic measurements, which would only provide the average diffusion values denoted by the distributions (≃ 1.2 cm$^2$.s$^{-1}$ and ≃ 1.4 cm$^2$.s$^{-1}$ as shown by a yellow dashed line in Figures 4c and 4d, respectively). These variations would also be missed using 1P PL measurements alone, which would only probe the surface. Therefore, these local, depth-dependent results further demonstrate the unique insight obtained by using the 2P microscopic technique.

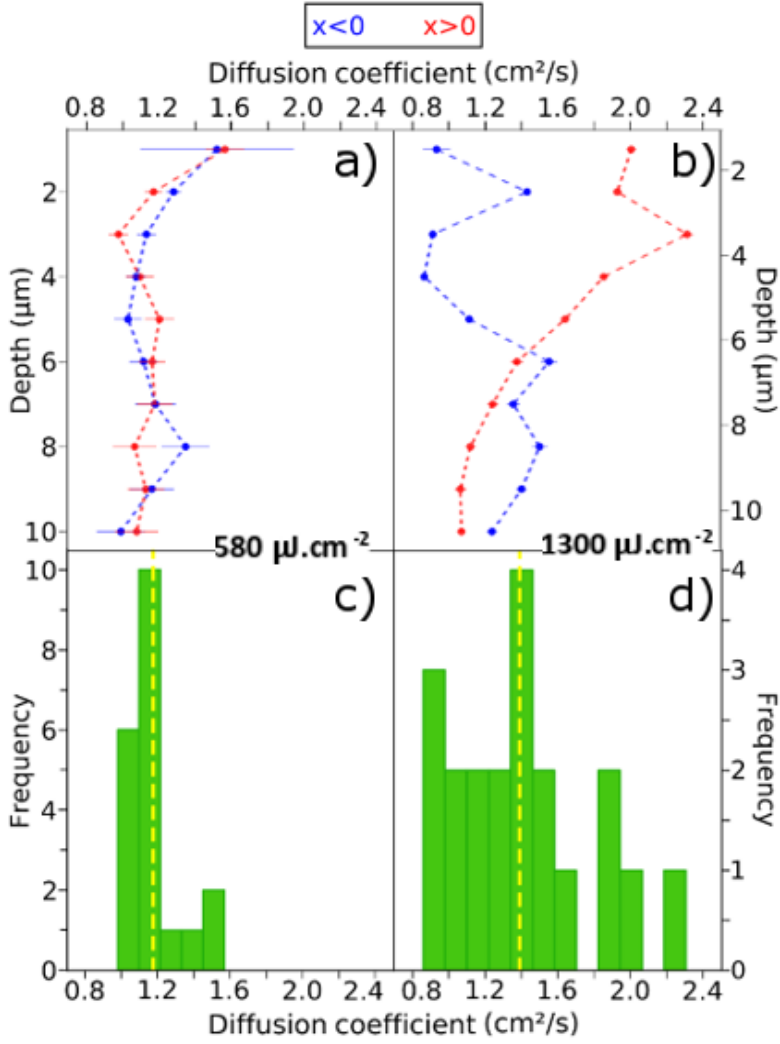

*Figure 4: Statistics of the depth-dependent diffusion coefficients in MAPbBr$_3$ single crystals. The depth-dependent (z) diffusion coefficients (D) obtained from fits to the diffusion plots in Figure 3 using Eq. 1, with excitation fluence of (a) 580 µJ.cm$^{-2}$ and (b) 1300 µJ.cm$^{-2}$. The regions x<0 (blue) and x>0 (red) are shown. The corresponding histograms of diffusion coefficients across all depths (z) and directions (x) are shown for the excitation fluence of (c) 580 µJ.cm$^{-2}$ and (d) 1300 µJ.cm$^{-2}$. The diffusion coefficients for the same z values are here binned together independently of the direction of carriers (x<0 or x>0). The dashed yellow lines denote the mean values of the distributions, which are ≃ 1.2 cm$^2$.s$^{-1}$ and ≃ 1.4 cm$^2$.s$^{-1}$, respectively.*

To better understand these heterogeneities, we display side-by-side in Figure 5 several important photophysical parameters obtained from the higher fluence (1300 µJ.cm$^{-2}$) 2P measurements for a range of spatial ($x$) and depth ($z$) values. The diffusion behaviour is highly asymmetric even below the surface, as large differences can be observed between the $x>0$ and $x<0$ profiles (Figure 5a). This is particularly evident between $z=2$ µm and $z=6$ µm (see yellow shaded area in Figure 5a and 5b), where we now focus our analysis. We observe that the diffusion coefficients are much larger for $x>0$ ($\simeq 2$ cm$^2$.s$^{-1}$) than for $x<0$ ($\simeq 1$ cm$^2$.s$^{-1}$). In Figure 5b, we show the PL decay time (defined as time taken for the PL to fall to 1/e of its initial intensity, see SI), averaged over the $x<0$ or $x>0$ lateral profile at each depth. We find that the PL decay time follows a very different trend than that of the diffusion coefficients, as the larger decay times are found on the x<0 side ($\simeq$ 4 - 8 ns) while the decay times for x>0 are appreciably shorter ($\simeq 2$ - 4 ns). In fact, the diffusion coefficients and PL decay times are anti-correlated in these two particular regions of the crystal. In Figure 5c, we show an $x$-$z$ image of the PL decay times (measured after excitation at $x=0$ for each depth). We see that the longer decay times for the $x<0$ region are measured over a region of several microns (inside the blue dashed circle region), extending in both $x$ and $z$ directions in that region. On the other side of the excitation region ($x>0$, red dashed circle), the decay times are comparatively lower and more spatially homogeneous. Additionally, the integrated PL intensity in the $x<0$ region (blue dashed circle) is a factor of 1.7 lower than the $x>0$ region (Figure 5d ; see Figure S11 for more details).

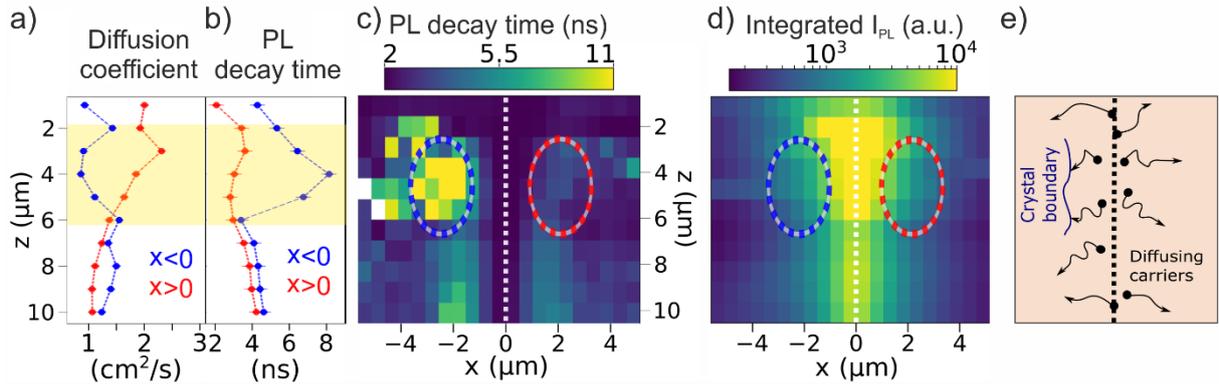

*Figure 5: Visualising a crystal boundary through photophysical measurements.* a) Diffusion coefficient and b) PL decay times (defined as the time taken to fall to 1/e of the initial intensity; see SI), averaged over the lateral profiles in each region at each depth, as a function of depth, as extracted from the data in Figure 3. The region *x<0* and *x>0* are denoted blue and red, respectively, and a region of interest is highlighted by yellow shading. x-z slices of the (c) PL decay time and (d) integrated PL intensity of the same region as in a) and b). Regions of interest discussed in the text are highlighted with blue (*x<0*) and red (*x>0*) dashed circles. e) Schematic showing the impact of a buried crystal boundary on the diffusion of carriers initially excited at *x=0 (dashed line).*

Given that there is a long PL lifetime but short diffusion coefficient and lower PL counts in the *x<0* region, we propose the presence of a defective crystal boundary between domains (Figure 5e) in the region in the blue dashed circle in Figure 5c and d. Indeed, edges and boundaries in halide perovskites crystals have been previously proposed to inhibit the diffusion of charge carriers[9]. Therefore, charge carriers moving through this x<0 area would be impeded from moving further beyond this boundary, leading to a lower effective diffusion coefficient in that region (see Figure 5e). Additionally, this model also explains why the increase in local carrier excitation density (fluence) has a negligible influence on the diffusion properties in this *x<0* region: such a physical barrier preventing the transport of charges may correspond to a defect type that is not able to be saturated in the same way as other point or extended defects, such as those in the *x>0* region. Indeed, boundaries often present a larger concentration of non-radiative recombination sites in halide perovskite materials[11,50], and their increased influence in that region may also explain the extended PL lifetime albeit lower PL intensity in that local region (lower fraction of radiative bimolecular recombination relative to non-radiative monomolecular processes[48]). Therefore, we conclude that charges near this boundary are significantly trapped, while the carriers at other depths are more freely able to diffuse (see Figure 5e).

In conclusion, we have developed a microscope platform to visualise in four dimensions (time and 3D space) carrier diffusion in different regions and depths of a semiconducting sample. We demonstrate its application on MAPbBr$_3$ single crystals, revealing local variations in charge-carrier diffusion on the microscale. At the surface, the diffusion is hindered by charge-carrier traps, but deeper in the sample we observe much larger diffusion coefficients that can even locally exceed the highest values reported in the literature from 1P TRPL measurements ($\simeq 1 cm^2.s^{-1}$ [29]). We use this technique to reveal a crystal boundary that impedes carrier diffusion even deeper into the crystal. This study demonstrates the capabilities of 2P TRPL tomography to visualise buried heterogeneities that would remain undetected with conventional 1P microscopy or macroscopic approaches. We expect the technique will be useful for a variety of semiconducting systems, ultimately providing guidance to improve the optoelectronic performance of devices.

**Experimental Section**

*Synthesis of the crystals*: MAPbBr$_3$ single crystals were prepared using an Inverse Temperature Crystallization method [42,43]. Specifically, a solid mixture of 0.672 g MABr (Dyesol Limited) and 2.202 g PbBr$_2$ (≥98%, Sigma-Aldrich) was dissolved in 4 ml DMF (anhydrous, 98%, Sigma-Aldrich) to form 1.5 M solution. This clear solution was filtered through 0.22-μm-pore-size PTFE filter and then divided into 5 vials. The vials were placed on a hot plate at room temperature and then slowly heated up to about 60°C at which the growth of MAPbBr$_3$ single crystals was achieved.

*Details of the 1P optical setup*: Confocal time-resolved one photon photoluminescence images and diffusion were measured using a confocal microscope setup (PicoQuant, MicroTime 200.) The excitation laser, a 405-nm pulsed diode (PDL 828-S"SEPIA II", PicoQuant, pulse width of around 100 ps), was directly focused onto the perovskite surface with an air objective (100x, 0.9 NA). The emission signal was separated from the excitation light (405 nm) using a dichroic mirror (Z405RDC, Chroma). The photoluminescence was then focused onto a SPAD detector for single-photon counting (time resolution of 100 ps) through a pinhole (50 μm), with an additional 410-nm longpass filter. Repetition rates of 10 MHz were used for the maps and the diffusion profiles. The lateral spatial resolution is ~550 nm.

*Details of the 2P optical setup*: For the 2P diffusion profiles, an optical fibre (25 μm core) was used for the raster scanning of the detection only. The collected photons were then sent onto a single-photon avalanche photodiode (SPAD), and their arrival times were recorded with a time-resolution of 100 ps. The 2P pulsed excitation is achieved using an Optical Parametric Oscillator (OPO, 150 fs pulses), set at a wavelength of 1200-nm (below the MAPbBr$_3$ bandgap) and a 100x air objective lens (NA = 0.95). The lateral spatial resolution of this optical microscope is estimated to be around 1.2μm in FWHM (and 0.5 μm in variance), the vertical spatial resolution is around 1.5 μm (see SI).

**Acknowledgements**

**§C.S. and G.D. contributed equally to this work**

A. A. Z. and O. M. B. gratefully acknowledge the funding support from King Abdullah University of Science and Technology (KAUST). Work at the Molecular Foundry was supported by the Office of Science, Office of Basic Energy Sciences, of the U.S. Department of Energy under Contract No. DE-AC02-05CH11231. C. S. thanks the EPSRC (Nano-Doctoral Training Centre), the Cambridge Trust and a Winton Graduate Exchange Scholarship for funding. This project has received funding from the European Research Council (ERC) under the European Union's Horizon 2020 research and innovation programme (grant agreement number 756962). S. D. S. acknowledges support from the Royal Society and Tata Group (UF150033). G.D. acknowledges the Royal Society for funding through a Newton International Fellowship. M.A. acknowledges funding from the European Union's Horizon 2020 research and innovation programme under the Marie Skłodowska-Curie grant agreement No.841386.

# Supporting Information

**Visualizing buried local carrier diffusion in halide perovskite crystals via two-photon microscopy**

Camille Stavrakas[§], Géraud Delport [§], Ayan A. Zhumekenov, Miguel Anaya, Rosemonde Chahbazian, Osman M. Bakr, Edward S. Barnard[c] and Samuel D. Stranks*

Dr. Camille Stavrakas, Dr. Géraud Delport, Dr. Miguel Anaya, Rosemonde Chahbazian, Dr. Samuel D. Stranks, Cavendish Laboratory, JJ Thomson Avenue, Cambridge CB3 0HE, UK.
*E-mail: sds65@cam.ac.uk

Ayan A. Zhumekenov, Dr. Osman M. Bakr, Division of Physical Sciences and Engineering, King Abdullah University of Science and Technology (KAUST), Thuwal 23955-6900, Kingdom of Saudi Arabia

Dr. Edward S. Barnard,
Molecular Foundry, Lawrence Berkeley National Laboratory, Berkeley, CA, USA

**Additional details of the one photon optical setup:**

For the 1P low fluence measurements, a power of 30 nW was used which corresponds to a fluence of 1.3 µJ.cm$^{-2}$. We can estimate the carrier concentration to be around $10^{17}$ cm$^{-3}$ (see discussion below) if we take into account the absorption depth of 100 nm in the present configuration[1] (405 nm excitation wavelentgh). For the high fluence measurements, a power of 300 nW was used which corresponds to a fluence of 13 µJ.cm$^{-2}$.

The raster scanning was performed using a galvo mirror system while both the objective and the sample remain at a fixed position. In the case of regular local TRPL measurements with this setup, both the excitation and the emission are scanned through the mirror system. On the other hand, only the emission path was scanned to create the diffusion profiles from the main text, while the excitation was decoupled and fixed at the center of the sample (*x=0*).

**Impact of the optical resolution to the width of the 1P TRPL beam:**

To evaluate the contribution of the optical resolution to the width of the PL Gaussian at *t=0*, we use the Abbe diffraction formula. In such a model, the distance between the maximum of the intensity and the first minimum is expressed as L=0.61*λ/NA, where NA=0.9 is the numerical aperture of the used objective lens. For the diffusion measurement, we evaluate the standard deviation σ of such intensity profile, under a Gaussian approximation of the profile. Therefore, the value of this variance can be estimated as $\sigma_{reso} \simeq 0.65 * L$. This yields the values of $\sigma_{reso} \simeq 180$ nm under excitation at 405 nm and $\simeq 240$ nm for emission at 540 nm.

**1P estimation of the TRPL decay:**

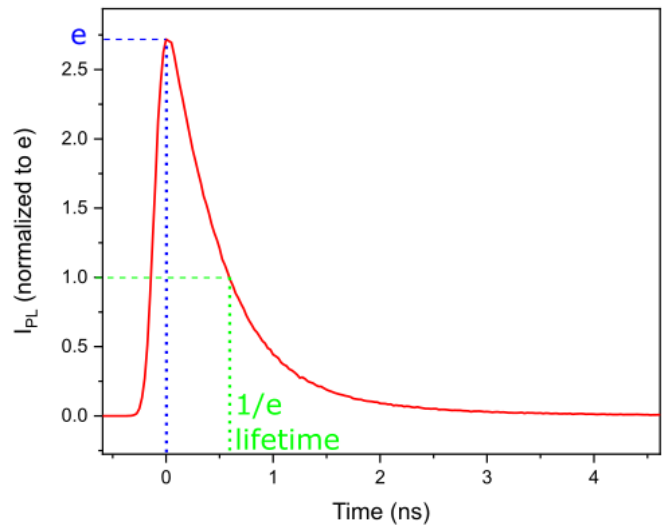

*Figure S1: 1/e Method to obtain the lifetime of the TRPL curve. The TRPL decay was estimated by measuring the delay at which the PL intensity has decreased a factor e from the initial intensity.*

**Series of 1P TRPL decays:**

x ≤0 part of the TRPL profile:

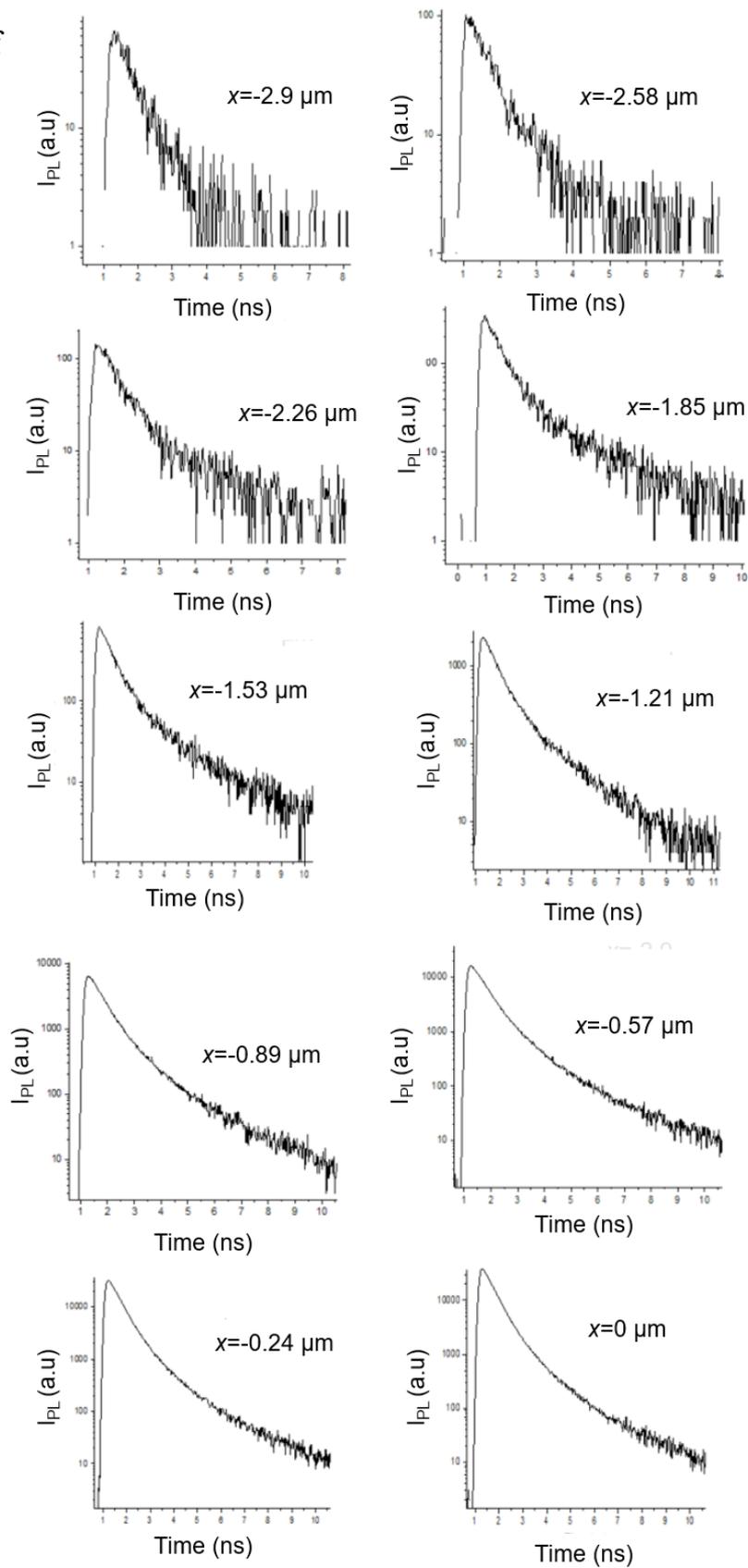

x≥0 part of the TRPL profile:

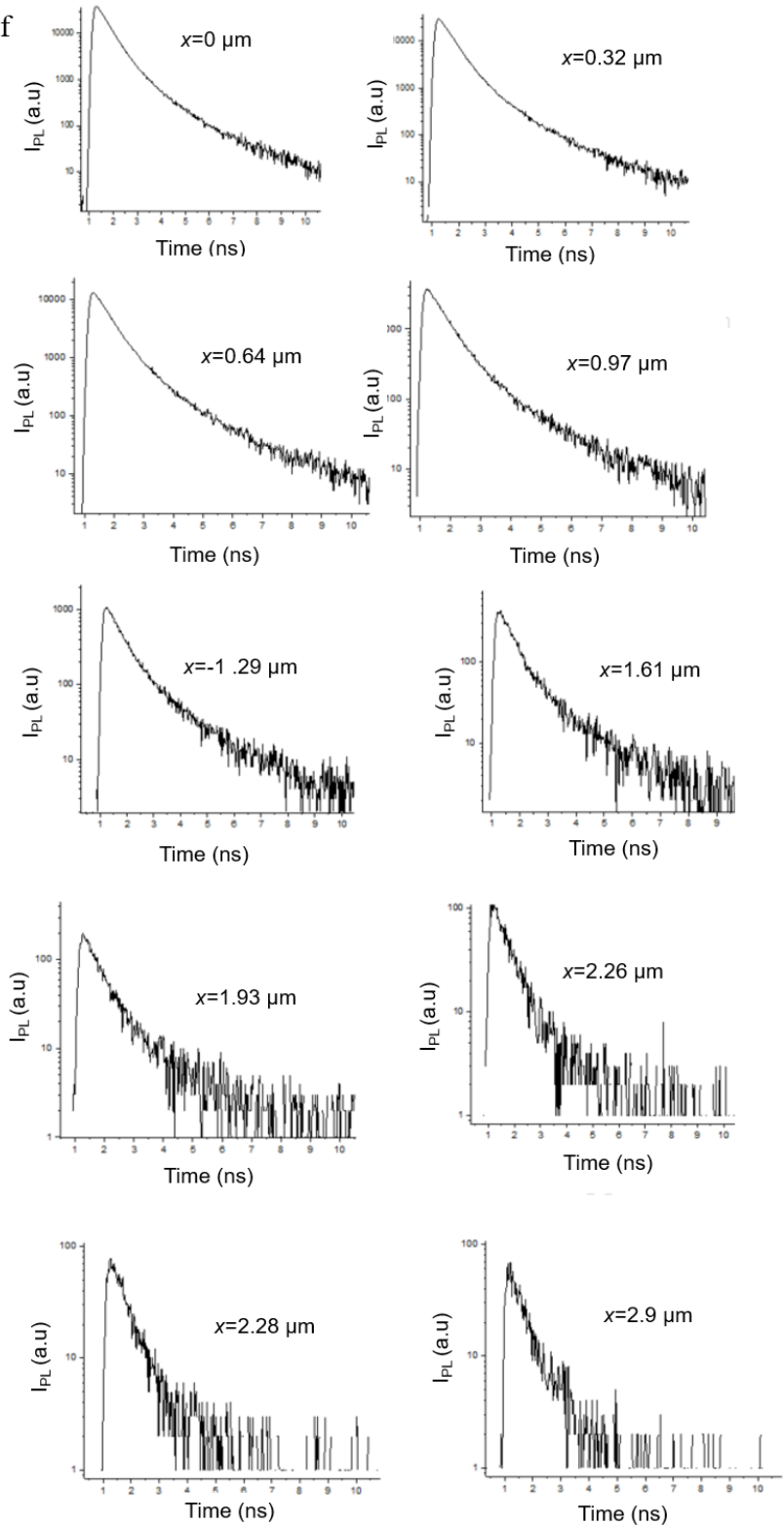

*Figure S2: two figures above: Series of TRPL decays obtained across the same 1P TRPL profile acquired by exciting at x=0 and collecting at different x values.* This forms two different sub-profiles for x<0 (above) and x>0 (below), that we use to calculate separately the diffusion coefficients on each side. The fluence is 1.3 µJ.cm$^{-2}$ and the repetition rate is 10 MHz.

**Additional 1P diffusion data:**

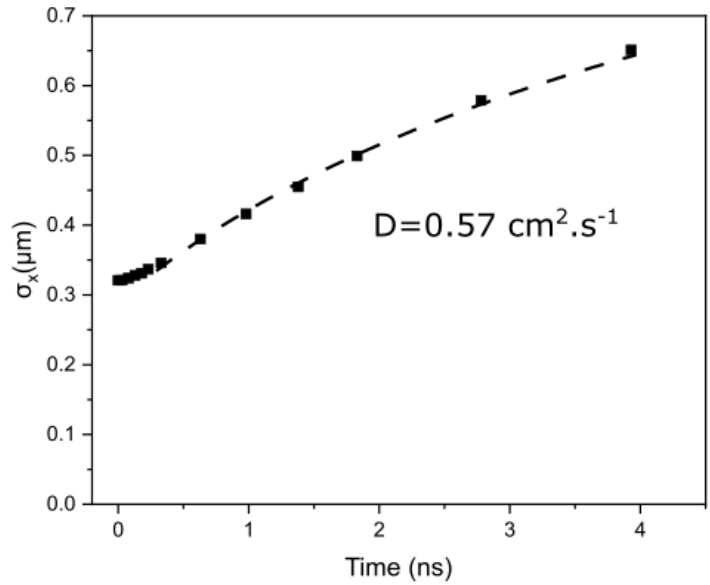

*Figure S3: Spreading of the PL beam as a function of time measured on a different point that the one displayed in the main text at a fluence of 1.3 µJ.cm⁻² . Here, only the x>0 spreading is displayed.*

**Comments on the light soaking effect:**

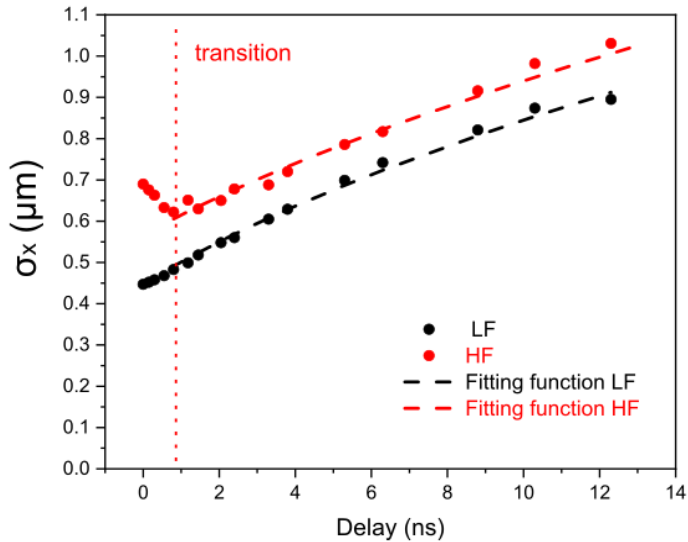

*Figure S4: 1P diffusion and light soaking measurements. a) Evolution of the variance as a function of time extracted from the TRPL diffusion profile for x<0 at a fluence of 1.3 µJ.cm⁻² (lower fluence, black) and 13 µJ.cm⁻² (higher fluence, red), including the diffusion law fitting (dashed lines). We observe a clear difference between the two profiles at early times, with an*

*artificial broadening of the initial variance at higher fluence due to the light soaking effect. However, the slope of the diffusive part is very similar. A comparable broadening effect is observed in the 2P measurements for z=1 µm.*

After a prolonged illumination of the sample, or when a higher fluence of 13 µJ.cm$^2$ is used, we observe a broadening of the initial PL width $\sigma_x(0)$ with respect to the lower fluence measurements. Additionally, $\sigma_x(t)$ stays almost constant on a timescale of around 1.5 ns. After this delay, it starts increasing with time indicating that the regular diffusive behaviour is progressively recovered. Interestingly, the value of the obtained diffusion coefficient in these conditions is 0.33 cm$^2$.s$^{-1}$, very similar to the one obtained without these intense excitation conditions. Therefore, the regular diffusion is still happening but another phenomenon induces a local change in the photophysical properties of the crystal. As an intense excitation is required to observe such effect, it may be connected to the light soaking effect. Interestingly, a previous study has reported a localized change in the PL properties of metal halide perovskites in the vicinity of the excitation spot, with similarities with our results[2]. Consequently, we conclude that such broadening is an indirect consequence of the diffusion of charge carriers and is likely connected to the lateral diffusion of ions under light soaking conditions changing the local material environment.

**Details of the 2P optical setup:**

Lateral steps of 500 nm have been used in the sample. By changing the depth-of-focus, we create this profile for different depths (z), with a step of 1 µm. On each location, the integration time for the total TRPL measurement is set to 20 seconds. For time-resolved collection, we spectrally-filtered the PL using a linear variable longpass filter to only get the red part of the spectrum and therefore minimize the influence of reabsorption and reemission on the measured diffusion. At each (x,z) coordinate, a 2P TRPL measurement was performed with a pulse repetition rate of ~8 MHz. We focus our study on a total time window of 7 ns that allows observing most of the intensity decay as well as the influence of the trapping dynamics on such decay. We used here two different fluences: a lower one of 584 µJ.cm$^{-2}$ and a larger one of 1300 µJ.cm$^{-2}$.

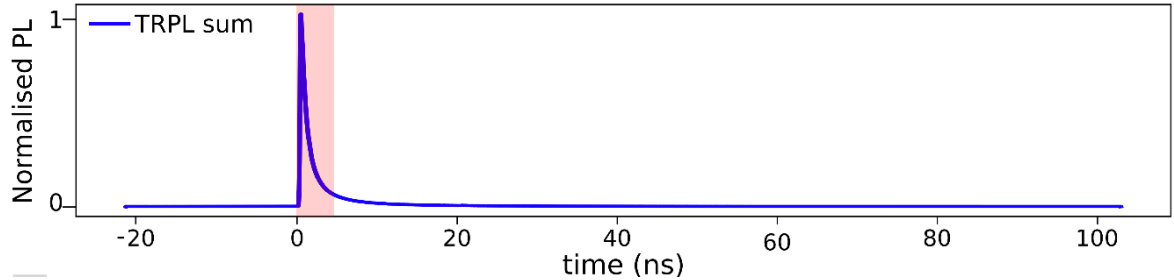

*Figure S5: Example of TRPL decay obtained with the 2P TRPL setup when exciting in the bulk of the crystal at the fluence of 1300 µJ.cm$^{-2}$. The red-shaded area highlights the time-interval used to construct the diffusion profiles.*

**Comparison of the 1P and 2P fluences:**

As mentioned above, the fluences used for the 1P TRPL measurements are either 1.3 µJ.cm$^{-2}$ or 13 µJ.cm$^{-2}$. This leads to respective estimated carrier concentrations of $2·10^{17}$ cm$^{-3}$ and $2·10^{18}$ cm$^{-3}$.

In the 2P configuration, the used fluences are 580 µJ.cm$^{-2}$ and 1300 µJ.cm$^{-2}$, with a pulse duration of 150 fs and an estimated beam vertical width of 1.5 µm. For these two fluences, we can estimate the pulse peak energy density to be respectively of 4 and 9 GW.cm$^{-2}$. Assuming a β coefficient of 8.6 cm·GW$^{-1}$ [3] we can use the formula published elsewhere[4] to estimate the photo-generated carrier concentrations in our measurements. For the two respective fluences, it yields $2.10^{17}$cm$^{-3}$ and $5.10^{17}$cm$^{-3}$, which are very similar values to the ones used in the 1P configuration. This allows us to compare the TRPL and diffusion results obtained in the 1P and 2P configurations.

**Series of 2P TRPL decays:**

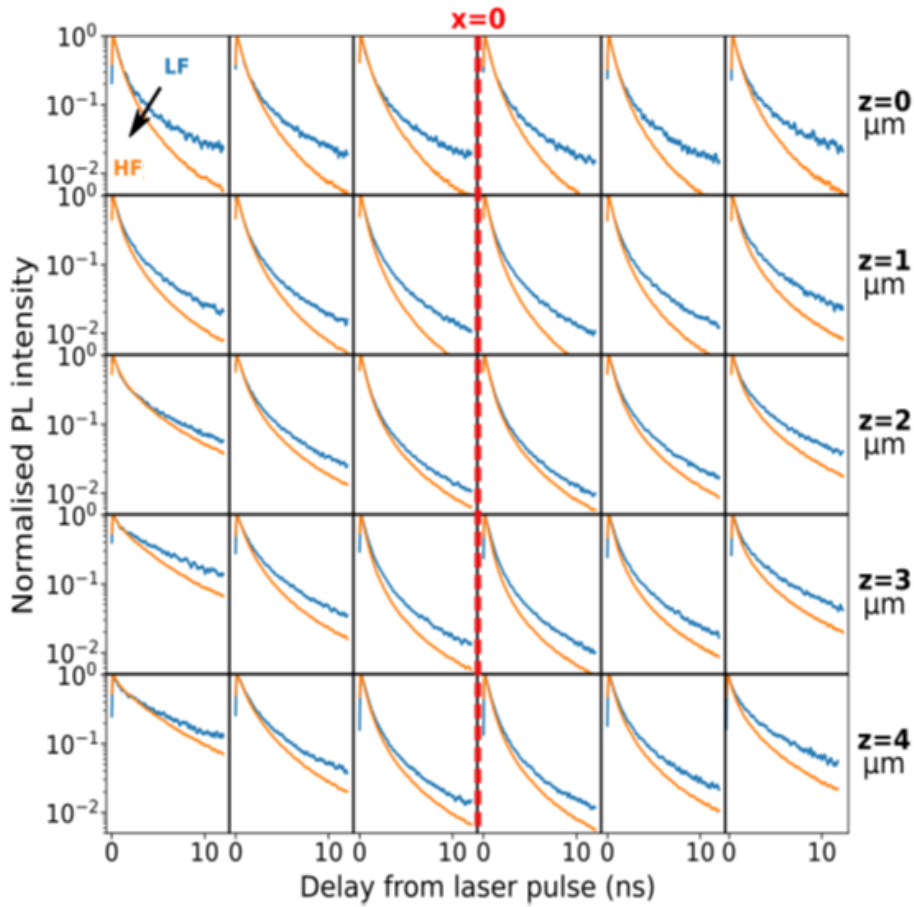

*Figure S6: Series of 2P TRPL decay curves for different x and z values on the MAPbBr3 crystal using 2P-TRPL at the two different fluences of 584 µJ.cm$^{-2}$ (lower fluence, LF, blue) and 1300 µJ.cm$^{-2}$ (higher fluence, HF, orange) highlighting the transition between a pure monomolecular to more slightly more bimolecular regimes.*

**Optical resolution of the 2P setup:**

In order to characterise the detection resolution of our 2P setup, we carry out a 2D (x,y) map of the laser reflection at the surface of a MAPbBr3 crystal. This map is shown as inset in Figure SI 1. We perform a Gaussian fit on the laser profile (Figure S7 and measure a $\sigma_{laser}$ ~ 0.5µm. This corresponds to a FWHM ~ 1.2µm, which is consistent with a diffraction-limited resolution.

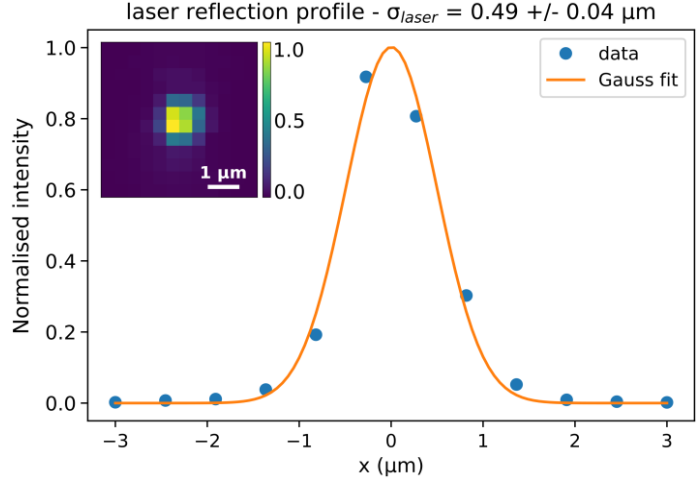

*Figure S7:* *The $\sigma_{laser}$ ~ 0.5μm of the laser reflection measured on the surface of a MAPbBr3 single crystal gives a lower bound to the imaging resolution with the raster-scanned collection fibre. The 2D (x,y) map of the laser reflection is shown as inset, the colour-scale representing the normalised light intensity.*

**TRPL decays and Accuracy of the Gaussian diffusion fittings.**

We fit each PL profile over a time window of 7 ns, highlighted in red in Figure S5. Figure S8 a-d shows a selection of these PL profiles and their respective fit as a function of depth and fluence at *t*=0 (Figure S8 a and b ) and *t*=7 ns (Figure S8 c-d). We note a good agreement of the data and the fit across the analysis window (0 ns to 7 ns). From the fit, we extract the time-dependent outward diffusion $\sigma_x(t)$. Figure S8 e and f are giving the mean error (as a percentage) on $\sigma_x(t)$ across the measurement window at low and high fluence, respectively. Contour lines help to visualise the confidence intervals and emphasize the large uncertainty (from 20% to >40%) arising at low fluence beyond *z*=6 μm after 3 ns. Despite the low signal-to-noise ratio in this quadrant, we are able to fit $\sigma^2_x(t)$ according to the diffusion equation (see main text) and obtain D values consistent with the rest of the dataset.

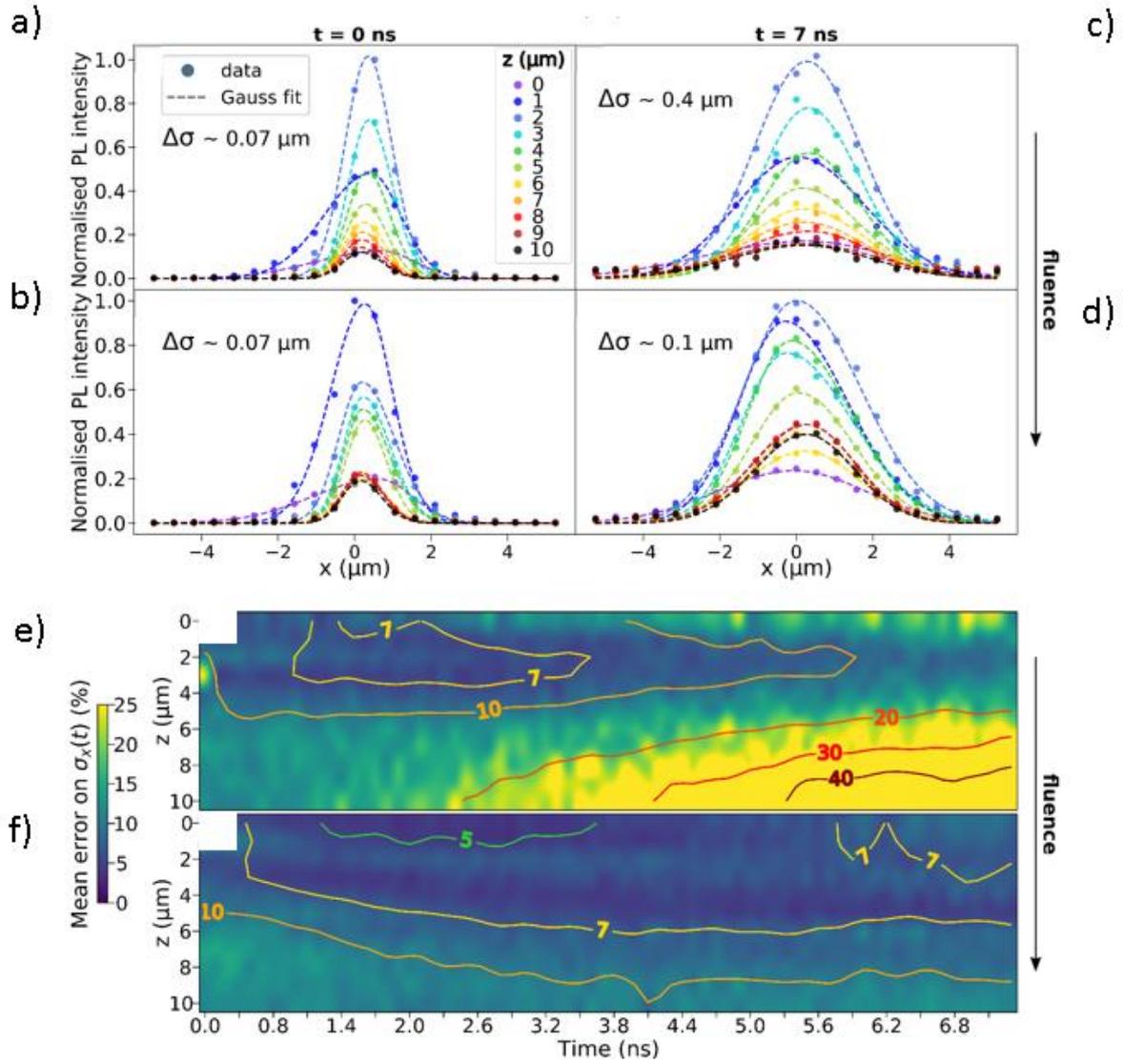

*Figure S8:* *Quality of the signal and analysis across the measurement window. PL profiles extracted from the 2P-TRPL measurement at different depths (z) and fluence for t =0 (a-b) and t = 7 (c-d) ns after the excitation pulse. An asymmetrical Gaussian fit is applied on each profile. Map of the error on $\sigma_x(t)$ extracted from the Gaussian fit, as a function of depth (z) and time, for lower (584 µJ.cm$^{-2}$, e) and higher (1300 µJ.cm$^{-2}$, f) excitation fluence. Contour lines help to visualise the variations in uncertainty from <5% to >40% of $\sigma_x(t)$.*

**Different diffusion profiles and fittings as a function of depth and fluence:**

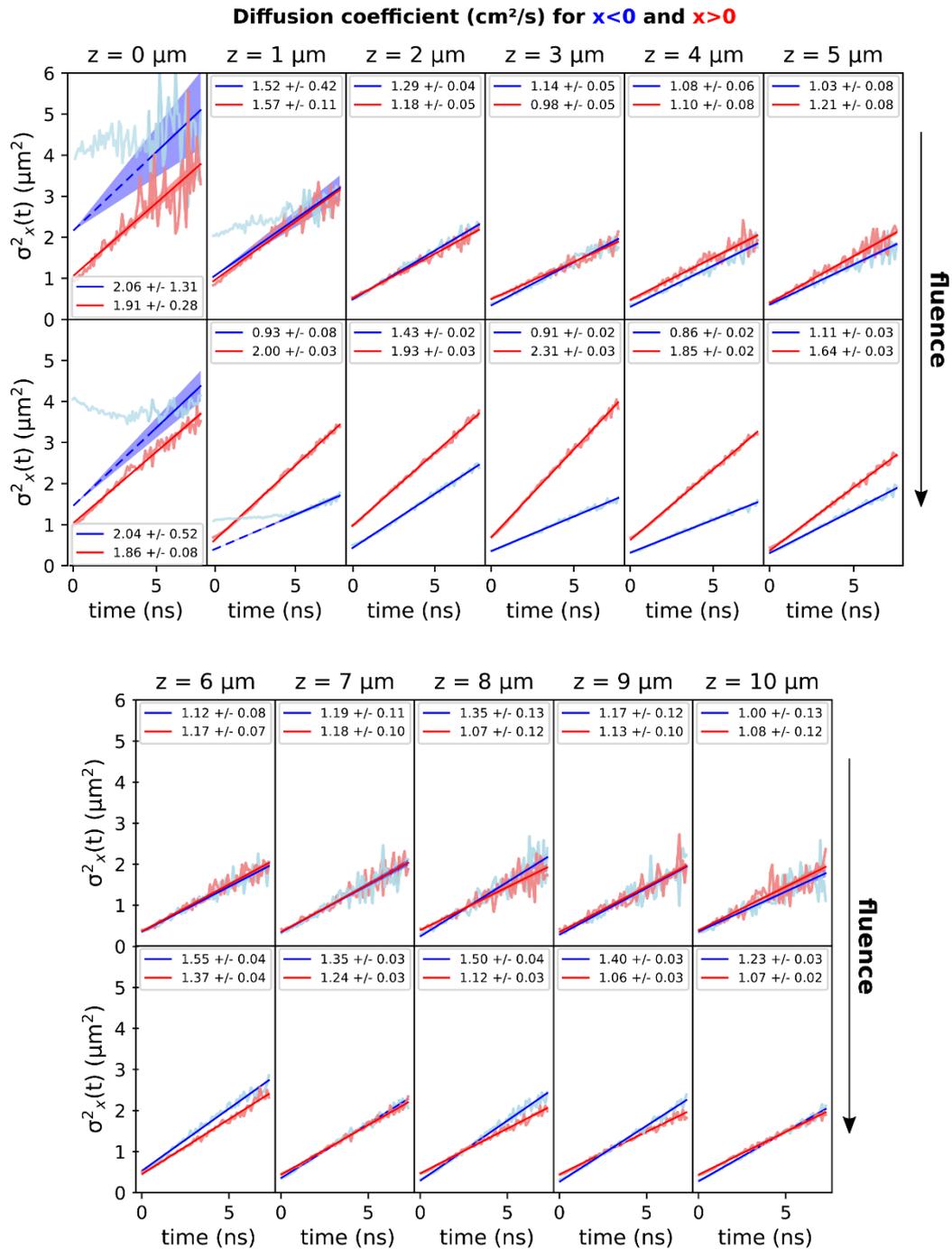

*Figure S9:* *Fitting of $\sigma^2_x$ (t) measured at different depths in a MAPbBr3 single crystal, with increasing fluence (584 µJ.cm$^{-2}$, top row and 1300 µJ.cm$^{-2}$, bottom row). The top and bottom figures show the data between z=0-5 µm and z=6-10 µm, respectively. The linear fit is overlaid on top of the data. Dashed lines indicate when an extrapolation was made. Filled areas around the fit help to visualise the standard error on the diffusion coefficient (slope).*

**Comparison of different photophysics parameter as a function of depth and fluence:**

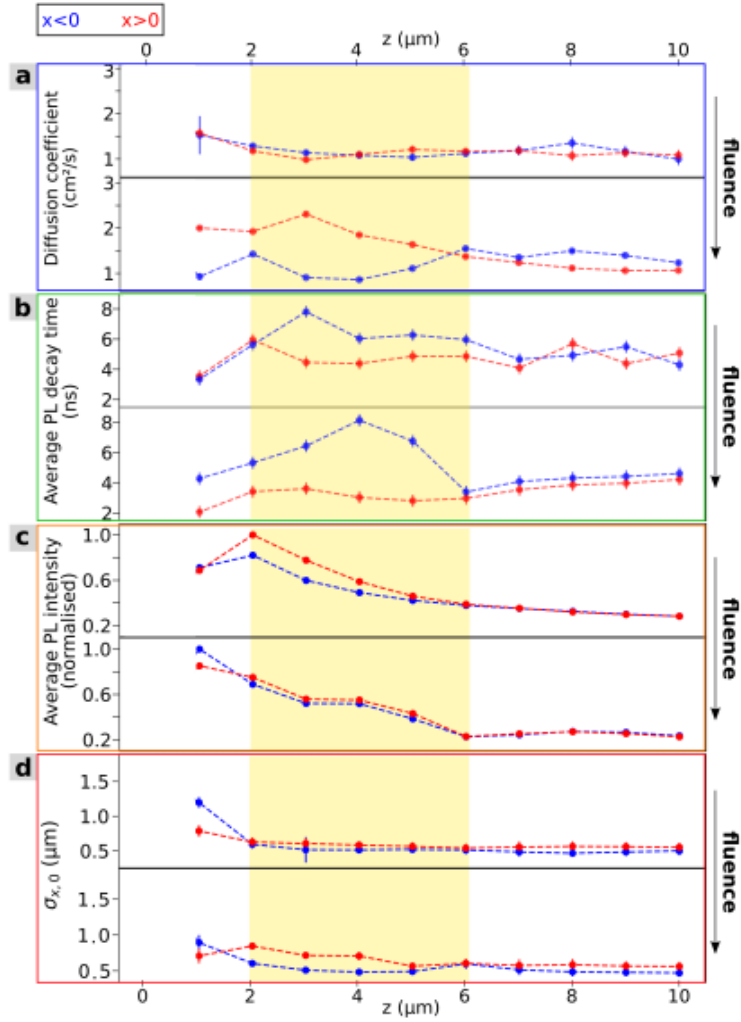

*Figure S10:* Direct comparison between our results in a MAPbBr3 single crystal. (a) the diffusion coefficient D, (b) the average PL decay time, (c) the average PL intensity and (d) the standard deviation a t=0 $\sigma_{x,0}$ are displayed as a function of depth and fluence in a MAPbBr3 single crystal. A yellow shaded window highlights the region where asymmetry is observed at high fluence, and where efficient diffusion and long decay times are anti-correlated. Used fluences were 584 μJ.cm$^{-2}$, *top row* and 1300 μJ.cm$^{-2}$, *bottom row*.

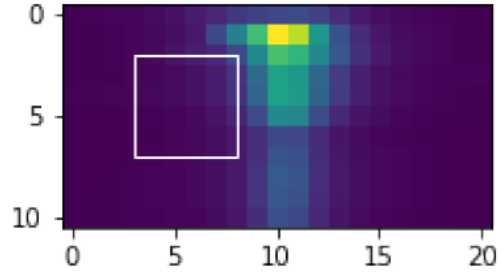
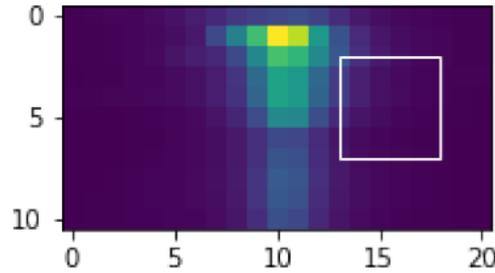

*Figure S11:* *The total count rate in the left rectangle (x<0) is 58% of the total count rate in the right rectangle (x>0). It shows that the hidden boundary on the left side leads to a quenching of the PL due a larger number of traps.*

**Derivation of the diffusion equation:**

Let us first look at the differential diffusion equation in 1D:

$$\frac{\partial c(x,t)}{\partial t} = D \frac{\partial^2 c(x,t)}{dx^2}$$

Where c is the physical quantity that diffuses as a function of time and D is the corresponding diffusion coefficient. In this article, c will be the carrier density.

In this equation, we have neglected the losses of carriers that mostly originates from the first order Shockley Read Hall [5] non-radiative recombination. As stated elsewhere[6], these losses will not affect the shape of the Gaussian beam can, therefore, be neglected as long as we consider the normalised distribution of carriers at every instant *t*. A particular solution above of this equation is of the form:

$$c(x,t) = \frac{1}{\sqrt{2\pi(2Dt)}} \exp\left\{-\frac{x^2}{2(2Dt)}\right\}.$$

This particular solution corresponds to an infinitely thin distribution of charge carriers for t=0 and x=0. At longer times, it corresponds to a Gaussian beam:

$$\sigma(t) = \sqrt{2Dt}$$

In the case of another initial distribution $c_0(x)$, the general solution can be written as the convolution product of the particular solution and the initial distribution of charge carriers:

$$c(x) = \int_{-\infty}^{-\infty} c_0(x-\xi) \frac{1}{\sqrt{4\pi Dt}} \exp\left\{\frac{-\xi^2}{4Dt}\right\} d\xi$$

In the case of a monochromatic (Laser) plane wave focused into a microscope objective, the initial distribution of carriers should be an Airy diffraction function. In the following, we will approximate this distribution with a Gaussian function of standard deviation $\sigma_0$:

$$c_0(x) = \frac{1}{\sigma_0 \sqrt{2\pi}} \exp\left\{\frac{-x^2}{2\sigma_0^2}\right\}$$

The full solution of the diffusion equation can be therefore expressed as:

$$c(x,t) = \frac{1}{\sigma_0 \sqrt{2\pi}} \times \frac{1}{\sqrt{4\pi Dt}} \int_{-\infty}^{-\infty} \exp\left\{\frac{-(x-\xi)^2}{2\sigma_0^2}\right\}) \exp\left\{\frac{-\xi^2}{4Dt}\right\} d\xi$$

which is the product of convolution of two Gaussian functions. Given the mathematical properties of the convolution product of two Gaussian functions, we know that distribution of charge carriers at each instant *t* will still be a Gaussian with a standard deviation $\sigma_1$ according to the next equation:

$$\sigma_1^2(t) = \sigma_0^2 + \sigma^2(t) = \sigma_0^2 + 2Dt$$

Due to the rotational symmetry of the studied system, this expression can be generalised to the 2D and 3D case by just considering each dimension as independent. As seen in Figure S12, a simple iterative simulation for an arbitrary Gaussian case in 1D, 2D and 3D confirms the relationship between the standard deviation and D, $\sigma^2(t) = \sigma^2_0 + ADt$ with $A = 2$. A simulation with a Gaussian squared, more relevant to the 2P excitation configuration, gives $A \sim 2.05$, very close to the Gaussian case.

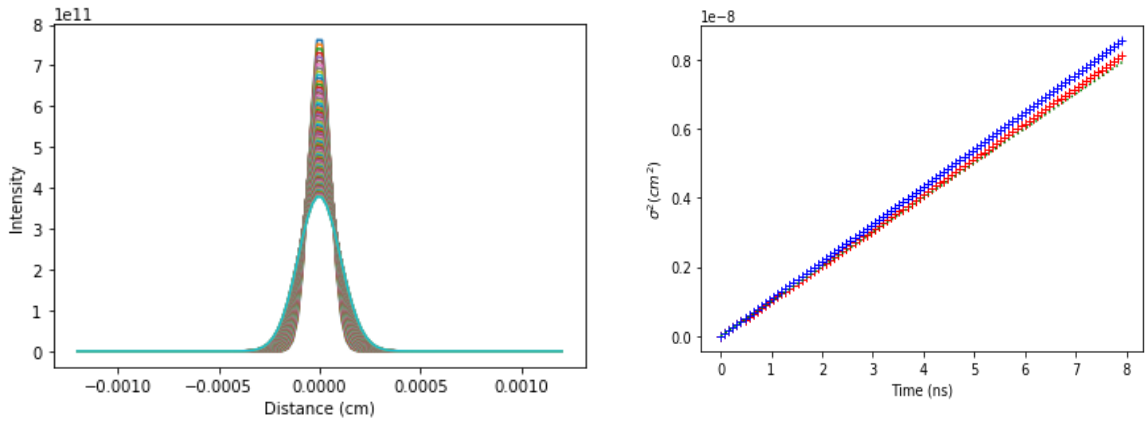

*Figure S12*: *Left: Evolution of the Gaussian beam at different times. Right: Plot of $\sigma^2(t) - \sigma^2(0)$ for a 1D (red crosses) and a 3D (blues crosses) configuration. The green crosses correspond to the theoretical result $\sigma^2(t) = \sigma^2_0 + 2Dt$.*

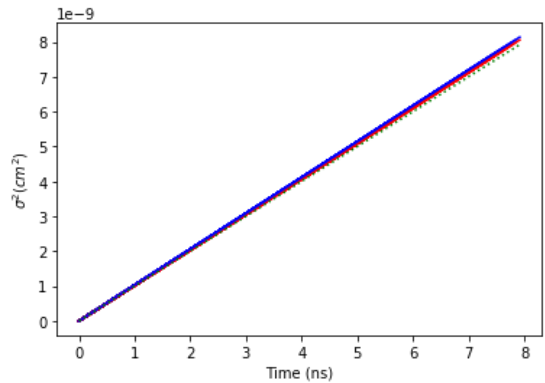

*Figure S13*: *Evolution of the variance $\sigma^2(t) - \sigma^2(0)$ for a 1D configuration with (blue) and without (red) the losses term. The green plot is the theoretical result $\sigma^2(t) = \sigma^2_0 + 2Dt$, showing a very small difference between the experimental estimation and the theory.*

Additionally, we have also included in our simulation a phenomenological first-order term in the diffusion equation to account for the loss of charge carriers due to the presence of traps:

$$\frac{\partial c}{\partial t} = D\frac{\partial^2 c}{\partial x^2} - \gamma_1 c$$

With our simulation, we have verified that this additional term does not modify significantly the evaluation of the diffusion coefficient with our Gaussian fitting process (see Figure SI 7). We have taken here $\gamma_1$=0.25 ns$^{-1}$.